\documentstyle[12pt]{article}
\begin{document}

\thispagestyle{empty}
\begin{flushright}{
          \parbox{5 cm}{
          \large UAB--FT--368\\
                 December 1995\\
                 astro-ph/9506040
          }}
\end{flushright}

\vspace{4cm}

\begin{center}
\begin{large}
\begin{bf}
Spin-Statistics Connexion, Neutrinos, and Big Bang
Nucleosynthesis
\end{bf}
\end{large}
\vspace{1cm}

\vspace{0.5cm}
L. Cucurull, J. A. Grifols and R. Toldr\`a \vspace{0.3cm}

IFAE. Grup de F\'{\i}sica Te\`orica.

Universitat Aut\`onoma de Barcelona

Bellaterra. Catalonia (Spain)
\vspace{2cm}

{\bf Abstract}
\end{center}

\begin{quotation}
We show how the $^4$He-abundance in the early Universe can be used to
demonstrate that macroscopic samples of neutrinos in thermal
equilibrium are indeed distributed according to Fermi-Dirac statistics.
\end{quotation}

\newpage

Cosmology is an excellent testing ground for theories and models of
Particle Physics ~\cite{kolb90}. Fundamental laws of Nature as well
ambitious
theoretical constructs have been scrutinized and confronted against
cosmological data. Conversely, Particle Physics provides important
clues to cosmological problems (e.g. provides candidates to dark
matter, provides explanations to primordial baryogenesis, etc.) A
paradigm of this fruitful symbiosis is the momentous prediction from Big
Bang Nucleosynthesis (BBN) on the number of light neutrino
species~\cite{steigman77}.

A fundamental result in relativistic quantum field theory is the
celebrated
spin-statistics theorem. Particles with half-integer spin obey
Fermi-Dirac statistics and particles with zero or integer spin obey
Bose-Einstein statistics. This theorem is deeply rooted  in very basic
principles such as relativistic invariance, locality and
microcausality and its experimental verification (and/or the
confirmation
of its consequences) is very important. Of course, the experimental
evidence of the spin-statistics connexion for ordinary matter in
macroscopic samples in overwhelming. Electrical (e.g.
superconductivity) and thermal (e.g. specifie heats) properties of
metals at low
temperatures can only be explained if electrons in matter obey
Fermi-Dirac statistics. Superfluidity too, is a reflection of the spin
properties of matter. The planckian blackbody spectrum of
radiation is just another phenomenological manifestation of the
spin-statistics theorem and traces back to the Bose-Einstein character
of statistical ensembles of photons.

Neutrinos, however, are  not usually found in macroscopic samples and
held in thermal equilibrium at a given temperature in a laboratory.
Thus, their statistical behaviour is difficult to be experimentally
established. Hence, a direct verification of the spin-statistics
connexion in the thermodynamic sense is still lacking in the case of
neutrinos. Sure enough, they are spin 1/2 particles, as a large amount
of data coming both from nuclear reactors and particle accelerators do
convincingly demonstrate. It would be nice then, as is the case for
other elementary constituents, to reveal the statistical mechanics of
neutrinos.

Two systems in Nature do contain a large macroscopic ensemble of
neutrinos in thermal equilibrium. A hot supernova core is one of them.
The other is the early Universe. In the cosmic system, one can
find a period in the early history
of the Universe where the spin-statistics connexion of neutrinos has
observational consequences. In the present paper we shall see that it
is possible to decide whether or not a macroscopic collection of
neutrinos in thermal equilibrium at temperature  T do indeed follow
Fermi-Dirac statistics. BBN and in particular the primordial helium
abundance $Y_P$ provides the basic tool for the analysis of this
question as we try to explain in the rest of this paper.

A fundamental
postulate of Quantum Mechanics is that systems of identical particles
are described by either symmetric or antisymmetric wave functions.
Systems of particles in thermal equilibrium whose states are described
by symmetric (antisymmetric) wavefunctions are distributed according to
Bose (Fermi) distribution functions. Now, it is a
phenomenological fact that electrons (and other particles with
half-integer spin) are described by anti-symmetric wave functions (and
obey Fermi-Dirac statistics when assembled  in large numbers and held
in thermal equilibrium) and photons (and other particles with integer
spin) are described by symmetric wavefunctions (and follow
Bose-Einstein statistics). That this should be so and not the other way
around is the content of the Pauli spin-statistics theorem, which
follows very generally from relativistic quantum field theory as
already mentioned above. The actual purpose of this investigation is to
realize that this is observationally the case for neutrinos.
So, our intention is not, to doubt about the fermionic nature of
neutrinos but rather, to explicitly display in a real physical system a
characteristic phenomenon associated to their nature. Namely, a large
collection of neutrinos in thermal equilibrium at temperature $T$ is
Fermi-Dirac distributed.

The helium-4 and other light element (i.e. deuterium, helium-3 and
lithium-7) primordial abundances have been periodically evaluated and
confronted with data embracing observations of ever increasing quality.
 The most recent analysis can be found  in references 3 to 9. The
theoretical calculation of the primordial helium abundance is by now
standard and the basic arguments can be read in the textbooks (see e.g.
ref. 10). The key quantity to be computed is the fraction $X_n$ of
neutrons to all nucleons as the Universe cools from equilibrium
temperatures well above neutrino freeze-out $(T>> 1 \ MeV)$ down to
the temperature where nucleosynthesis takes place $(T \sim 0.1 \ MeV)$.
$X_n$ is found from the evolution equation
\begin{equation}
\frac {dX_n}{dT} \frac {dT}{dt} = - \lambda (n \to p) X_n + \lambda (p
\to n) (1-X_n)
\end{equation}

where $\lambda (p \leftrightarrow n)$ are the weak rates (per nucleon)
that interconvert protons and neutrons. These rates contain the
microphysics (weak interaction probabilities) and the macrophysics
(thermalization). It is this latter component that we shall manipulate.

The general strategy to test the statistical character of neutrinos in
the cosmic sample will be to use Bose-Einstein statistics as an
alternative templet in order to check how dependent the helium
abundance
actually is on the Fermi-Dirac distribution functions (it could well be
that the choice of statistics were irrelevant). We do not consider
here other kinds of statistics (e.g. parastatistics) which could also
lead to  sound field theories. An other place where the choice of
statistics matters -apart from the rates
$ \lambda(n \leftrightarrow p)$ above-
is the law $T=T(t)$ that enters eq. (1) since the expansion of the
Universe depends on the fermionic/bosonic effective degrees of freedom.
Here we choose {\underbar {all}} three neutrino species (in the rates
only the electron neutrino enters) to obey Bose statistics.

In order to contrast the Fermi behaviour against an alternative
behaviour one should
solve eq. (1) using neutrinos with the right (F-D) statistics and find
also the neutron  fraction in the case of
neutrinos with the wrong (B-E) statistics, i.e. we replace everywhere
$(e^{E/\kappa T_{\nu}}+1)^{-1}$ by $(e^{E/\kappa T_{\nu}}-1)^{-1}$ and
replace Pauli
blocking factors $1-f_{FD}$ for neutrinos wherever they appear
in the conventional
calculation (see ref. 10) by stimulated emission factors $1+f_{BE}$ .
One may check at this point the self-consistency of using B-E statistics
for neutrinos by proving that, at high temperature, detailed balance is
satisfied, i.e. $\lambda(n \to p)= \lambda(p \to n),$ and hence
equilibrium can be maintained. Indeed, for $(m_n - m_p)/\kappa T << 1,$
with $T=T_\gamma=T_\nu$,
\begin{equation}
\lambda (n \to p) = \lambda (p \to n) =
const \times \int^\infty_{-\infty} dq \
q^4
f (q/ \kappa T)
\end{equation}
where
\[ f(x) \equiv \pm (1+e^{-x})^{-1} (1-e^x)^{-1} \quad \]
the upper (lower) sign for negative (positive) $x$.

We have implemented the above modifications in Kawano's version [11] of
the BBN code by Wagoner [12] to obtain the relative effect on the
primordial abundances of $^4He,\  D,\ ^3He$ and $^7Li$ induced by the
change in statistics. We define the quantities,
\begin{equation}
\frac{\Delta a_i}{a_i} \equiv \frac{a_i(BE)-a_i(FD)}{a_i(FD)} \quad
i=1,2,3,4
\end{equation}
where
$a_1 \equiv Y_P, \ a_2 \equiv [D/H]_P, a_3 \equiv [^3He/H]_P$ and
$a_4 \equiv [^7Li/H]_P$. Our results are shown in figure 1.

Being the number of light neutrino species no longer a free parameter
(but rather, fixed by LEP data to be 3) the only free parameter in our
BBN calculation in $\eta_{10}$, related to
the baryon density $\Omega_B$ of the
Universe  through
\begin{equation}
 \Omega_B = 0.0036 h^{-2} (T/2.726)^3 \eta_{10}
\end{equation}
where $T$ is the microwave background temperature today and $h$ defines
the Hubble parameter $H=100 h \ km \sec^{-1}/Mpc$. Therefore,
figure 1 shows
the quantitites $\Delta a_i/a_i$ as a function of $\eta_{10}$. The
effect on the helium abundance $Y_P$ -the one we are primarily
interested in since it is the one which is more severely constrained by
observation- is negative and almost 4\%. This number is the net result
of two competing sources: a change in the weak interaction rates and a
change in the expansion rate. The variation due to the the Hubble
expansion is necessarily positive because trading fermionic degrees of
freedom for bosonic degrees of freedom implies an acceleration of the
expansion rate and, as a consequence a larger amount of $^4He$
produced.
In fact, it amounts to $+2.3\%$ relative increase as we have explicitly
checked. The dominating effect lies therefore in the $\lambda
(n\leftrightarrow p)$ rates which turns out to be negative.

We are now prepared to confront our results with observation.
The basic observational quantity is $Y_P$ and we shall use
the primordial abundances of other light elements $(D, ^3He$ and
$^7Li$) to help us fixing a value or range of values for  $\eta_{10}$.
The recent detection of the Lyman $\alpha$ line of deuterium in a Quasar
Absortion System (QAS) at high redshift (z=3.32) can be interpreted as
determining the primordial deuterium abundance:~\cite{songaila94,
carswell94}
\begin{equation}
[D/H]_P=(1.9 -2.5) \times 10^{-4}
\end{equation}
It is an order of magnitude larger than the measurement by the Hubble
Space Telescope of the D abundance in the local interstellar medium
(LISM)~\cite{linsky93}
\begin{equation}
[D/H]_{\rm{LISM}}=(1.6^{+0.07}_{-0.18}) \times 10^{-5}
\end{equation}
Although strictly speaking there is no contradiction between eqs. (5)
and (6),  since eq. (6) has to be interpreted as a lower bound on the
primordial D abundance
\begin{equation}
[D/H]_P \geq 1.5 \times 10^{-5},
\end{equation}
nonetheless they cannot be made to agree when extrapolating eq. (6)
deep into the past by using galactic chemical evolution models.

If one takes eq. (5) seriously then a very good overall agreement
between theory and data can be achieved with a value of $\eta_{10}=1.6
\pm 0.1$ derived from the matching of BBN and the QAS deuterium
abundance measurement~\cite{dar95}.
The dependence of $[D/H]_P$ on $\eta_{10}$ is much stronger than it is
the case for $Y_P$ and, hence, small variations on deuterium abundance
associated to the change in statistics as the ones shown in fig. 1, do
not lead to appreciable modifications in the determination of
$\eta_{10}$. Therefore, we can use the same $\eta_{10}$ range in both,
Fermi-Dirac and Bose-Einstein cases.

The above determination of $\eta_{10}$ in turn implies a band of values
for the helium abundance $Y_P$. We display our result in fig. 2.
In the figure we also collect
data associated to recent analysis of observations of the helium
abundance. The errors shown are 1 $\sigma$ errors and we refer to the
original work [ refs. 3-7] for a full discussion of their nature and for
an appraisal of the techniques of analysis involved. We
also include a weighted average (full circle). The gray shaded column
on the right is the standard BBN prediction and the darker column
on the left is our result for neutrinos with
the wrong statistics (the width of the bands reflects both the spread
in $\eta_{10}$ and the experimental uncertainty in the neutron half
life: $\tau_n =887 \pm 2.0 \ sec$ ~\cite{particle94}).
To 1  $\sigma$, the spin-statistics connexion for
neutrinos is confirmed (i.e. we can exclude the alternative templet).

However, we should investigate the dependence of our statement on the
chosen value of $\eta_{10}$. As already mentioned above, other sets of
data suggest lower deuterium abundances, which in turn mean higher
values for $\eta_{10}$. A reasonable window in the intermediate segment
of $\eta_{10}$ values should be 2.5 $\leq \eta_{10} \leq$ 3. In the high
end of $\eta_{10}$ we may take $\eta_{10}$=6 which is
comfortably larger than the
upper limit $\eta_{10}=5.27$ derived from matching primordial $^7Li$
abundance and BBN~\cite{kernan94}.
Figure 3 shows the low, intermediate and high domains of $\eta_{10}$.
The vertical dotted lines are the 1 $\sigma$ limits on the observed
(average) helium abundance (the fat error bar in fig. 2). The gray and
dark-gray boxes correspond to standard BBN results ("right" statistics)
and to the results of our analysis (neutrinos with "wrong" statistics),
respectively. A comment on the interpretation of this figure may
serve as a concluding summary on the content of the paper.

First, note that the low part of fig. 3 is just a restatement of
fig. 2. It best exemplifies the purpose of this work, i.e. to show that
BBN has the potential to decide empirically whether neutrinos verify
the spin-statistics theorem. The middle part of the figure tells us
that B-E statistics (the alternative templet) cannot be ruled out
which, of course, means that the F-D statistics of neutrinos
cannot be demonstrated. It shows
also that, to 1$\sigma$, even standard BBN can be in trouble.
Better data are
therefore required. Finally, the top part of fig. 3 informs us that
standard BBN
does not work (even at the 2$\sigma$ level) and hence the statistics
issue is irrelevant in this case.

A last comment is in order. No matter what the final fate of the
controversial deuterium data [13,14] shall be, the avalanche of new and
better data in the coming years will certainly permit a positive
discrimination of neutrino statistics. As an illustration of this
point, we note that a modest 20\% reduction of the error bars  in the
helium data suffices to make the distance between boxes in fig. 3
larger then the 1-$\sigma$ error interval.

\bigskip
\noindent{\Large \bf{Acknowledgements}}

\bigskip

Work partially supported by CICYT under project AEN-93-0474 and
AEN-95-0882. We thank
the Theoretical Astroparticle Network for support under the EEC
Contract No. CHRX-CT93-0120 (Direction Generale 12 COMA).
J.A.G. thanks G.
Raffelt for a helpful conversation. R.T. acknowledges a FPI Grant from
the Ministerio de Educaci\'on y Ciencia (Spain). We thank also
Prof. E. Kolb for
kindly providing us with the BBN code cited in the text.

\newpage

\noindent{\large \bf Figure Captions}

\bigskip

\noindent {\bf Fig. 1.} The various quantities $\Delta a_i/a_i$ as
defined in the text, as a function of $\eta_{10}$.
\bigskip

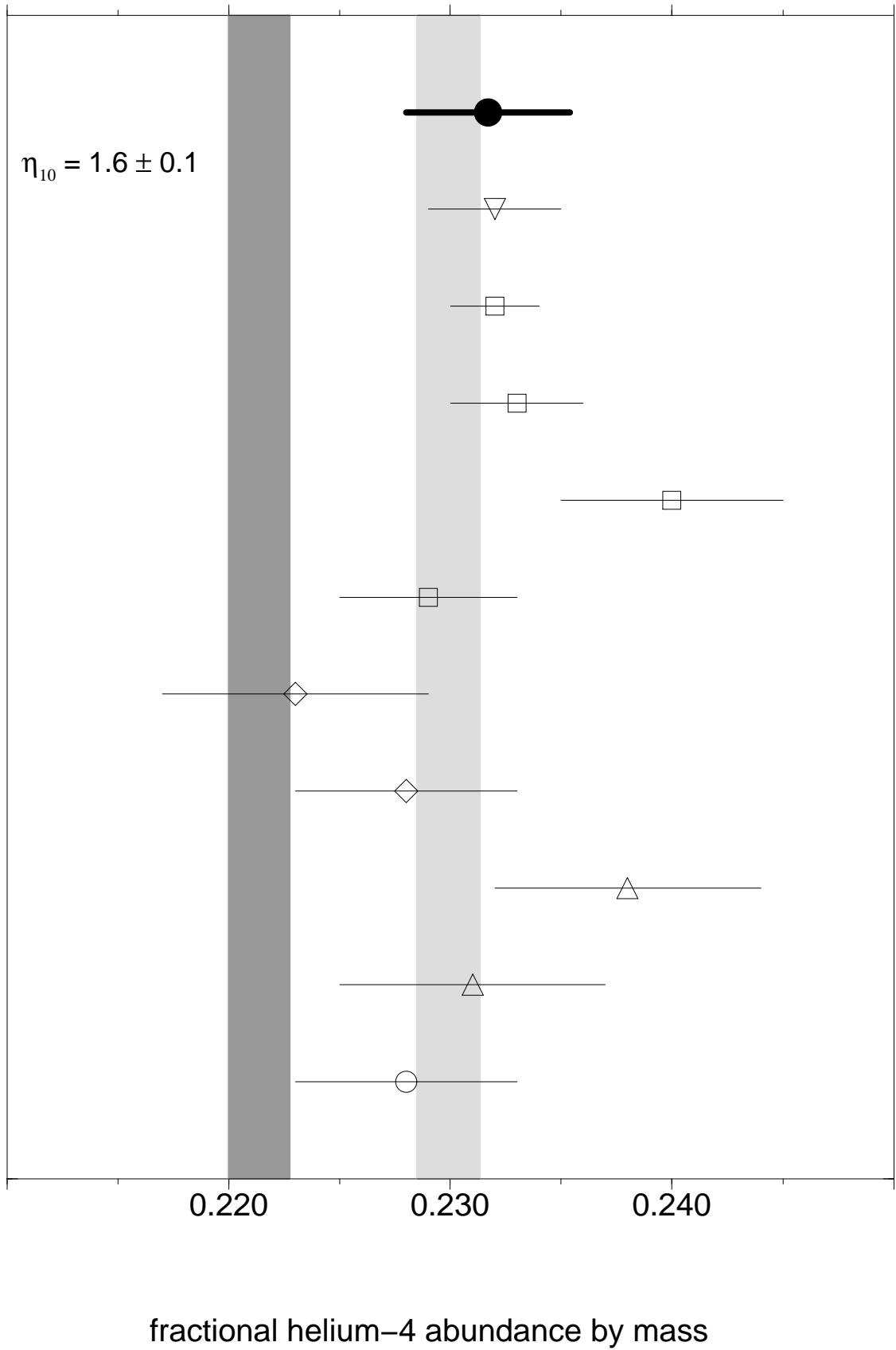
\noindent {\bf Fig. 2.} Confronting F-D statistics (gray band) and B-E
statistics (dark-gray band) with He-4 data from: ref. 3
(circle), ref. 4 (diamonds) ref. 5 (triangles), ref. 6 (squares), ref. 7
(inverted triangle). The full circle is the average value. Here,
$\eta_{10}=1.6 \pm 0.1$.

\bigskip
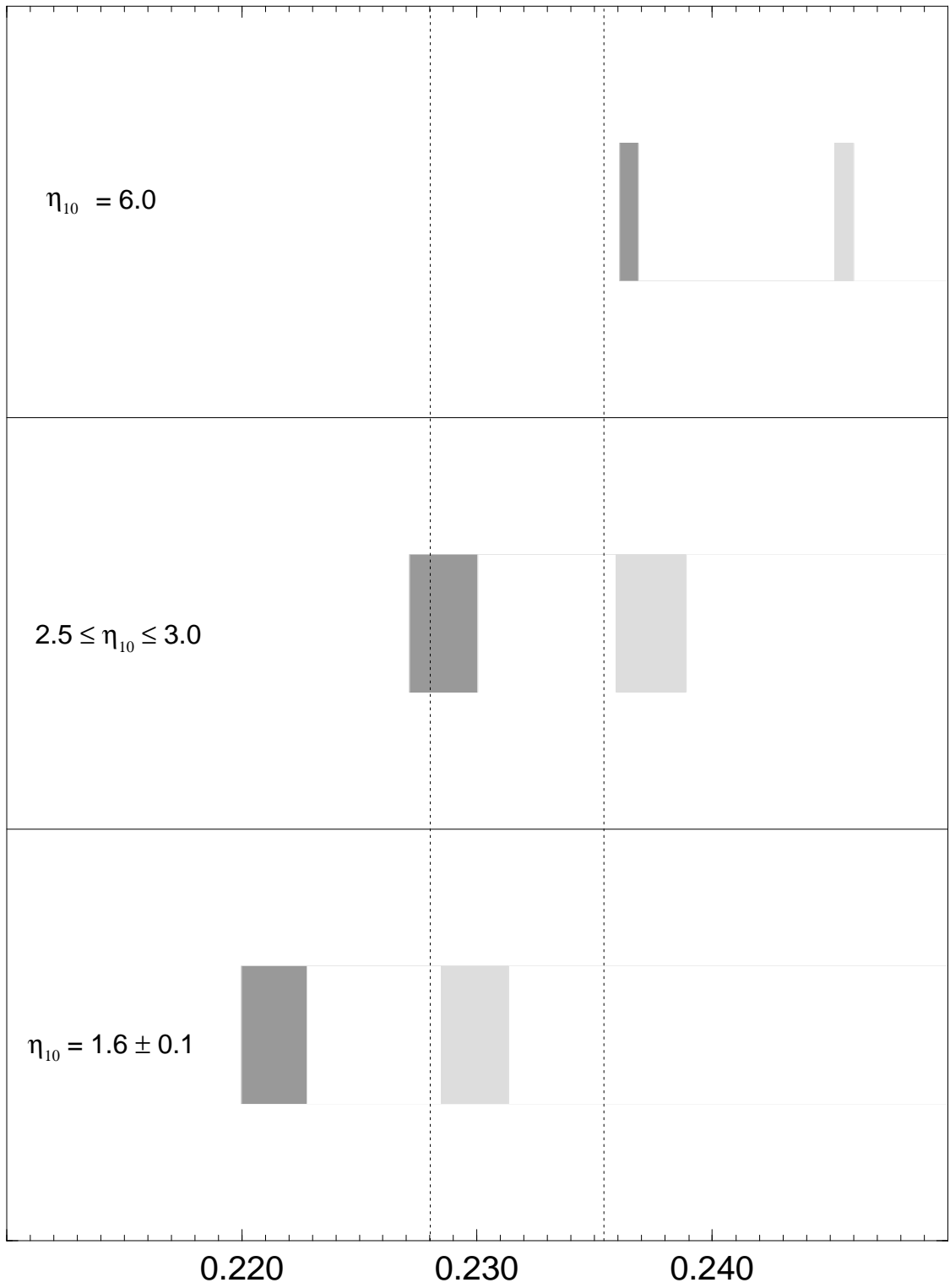
\noindent {\bf Fig. 3.} Our analysis displayed for three domains of
$\eta_{10}$. The two vertical dotted lines bracket the observationally
allowed He-4 abundances as given by the average abundance shown in fig.
2.

\end{document}